\documentstyle[prl,aps,twocolumn,epsf]{revtex} 

\begin{document}
\draft
\wideabs{

\title{On the theory of Josephson effect in a diffusive tunnel junction}

\author{E.V. Bezuglyi, E.N. Bratus', and V.P. Galaiko}

\address{B. Verkin Institute for Low Temperature Physics and Engineering, \\
National Academy of Sciences of the Ukraine, 310164 Kharkov, Ukraine
\\E-mail: bezugly@ilt.kharkov.ua}

\date{Submitted September 16, 1998}

\maketitle

\begin{abstract}
Specific features of the equilibrium current-carrying state of a Josephson 
tunnel junction between diffusive superconductors (with the electron mean 
free path $l$ smaller than the coherence length $\xi_0$) are studied 
theoretically in the 1D geometry when the current does not spread in the 
junction banks. It is shown that the concept of weak link with the jump 
$\Phi\sim 1$ of the order parameter phase exists only for a low transmissivity 
of the barrier $\Gamma \lesssim \l/ \xi_0 \ll 1$. Otherwise, the presence of 
the tunnel junction virtually does not affect the distributions of the order 
parameter modulus and phase. It is found that the Josephson current induces 
localized states of electron excitations in the vicinity of the tunnel 
barrier, which are a continuous analog of Andreev levels in a ballistic 
junction. The depth of the corresponding ``potential well'' is much greater 
than the separation between an Andreev level and the continuous energy
spectrum boundary for the same transmissivity of the barrier. In contrast to 
a ballistic junction in which the Josephson current is transported completely 
by localized excitations, the contribution to current in a diffusive junction 
comes from whole spectral region near the energy gap boundary, where the 
density of states differs considerably from its unperturbed value. The 
correction to the Josephson current $j(\Phi)$ in the second order of the 
barrier transmissivity, which contains the second harmonic of the phase jump 
$\Phi$, is calculated and it is found that the true expansion parameter 
of the perturbation theory for a diffusive junction is not the tunneling 
probability $\Gamma$, but a much larger parameter $W = (3\xi_0/4l)\Gamma$. 
This simplifies the conditions for the experimental observation of higher 
harmonics of $j(\Phi)$ in junctions with controllable transmissivity of the 
barrier.
\end{abstract}
\pacs{Fiz. Nizk. Temp. {\bf 25}, 230--239 (March 1999)}
}
\narrowtext

\section{Introduction}

In recent years, considerable advances have been made in technology of 
preparing low-resistance tunnel junctions with a comparatively high barrier 
transmissivity (tunneling probability) $\Gamma$. This primarily applies to 
controlled break-junctions \cite{1} as well as systems based on 2D electron 
gas \cite{2}, whose conductivity undergoes a crossover from tunnel to metal 
type upon a change in the barrier parameters. The problem of calculation of 
the Josephson current through a junction with an arbitrary transmissivity in 
the ballistic regime (with the electron mean free path $l$ much greater than 
the coherence length $\xi_0$) was solved by many authors \cite{3} on the 
basis of the model of a single-mode junction with current-carrying banks 
ensuring a rapid ``spreading'' of supercurrent and the equality of the order 
parameter modulus $\Delta$ near the barrier to its bulk value (the 
``rigidity'' condition for $\Delta$).

In the 1D geometry (e.g., a planar junction or a superconducting channel with 
a tunnel barrier \cite{note1}), the problem is complicated considerably due 
to the change in the order parameter and the quasiparticle energy spectrum in 
the vicinity of the junction, which makes a contribution to the phase 
dependence of the current $j(\Phi)$. Antsygina and Svidzinskii \cite{4} 
determined the corresponding corrections to $j(\Phi)$ of the order of
$\Gamma^2$ for a pure ($l \gg \xi_0$) superconductor in the limit of low 
transmissivity $\Gamma \ll 1$:
\begin{equation}
\delta j(\Phi)\! =\! -\alpha(T)I(\Delta)\Gamma \left(\sin \Phi\! -
\!{1 \over 2}\sin 2\Phi\right)\!, \;\; \alpha(T)\! \sim\! 1,
\end{equation}
\begin{equation}
I(\Delta) = (\pi/ 4)e\nu_F v_F \Gamma\Delta = I_c(\Delta)/\tanh(\Delta/2T),
\end{equation}
where $\nu_F$ is the density of states, $v_F$ the Fermi velocity, and 
$I_c(\Delta)$ the critical current through the junction.

In a diffusive superconductor (the ``dirty'' limit $l \ll \xi_0 =
\sqrt{D/2\Delta}$, $D = v_Fl/3$ is the diffusion constant), the calculation 
of the Josephson current for an arbitrary $\Gamma$ is hardly possible 
\cite{note2} even in a simple model disregarding the variation of the order 
parameter in the vicinity of the junction. As a matter of fact, the 
boundary conditions for isotropic Green's functions $\hat{g}(\bbox{r},t_1t_2)$ 
at the junction, obtained by Kupriyanov and Lukichev \cite{7}, 
\begin{equation}
-l(\hat{g}\nabla\hat{g})_L = -l(\hat{g}\nabla\hat{g})_R = 
{3\over 4}\left<{\mu d(\mu)\over r(\mu)}\right>
\left[\hat{g}_L, \hat{g}_R\right], 
\end{equation}
%
where $d(\cos\theta)$ is the tunneling probability for an electron impinging 
the barrier at an angle $\theta$, and the subscripts $R$ and $L$ mark the 
value to the right and left of the barrier, are valid only within the first 
order in small angle-averaged transmissivity $\Gamma = \langle \mu d(\mu)
\rangle$. Lambert et al. \cite{8} proved that the derivation of the boundary 
conditions in the general case ($d \lesssim 1$) is reduced to an analysis of 
a system of nonlinear integral equations for the terms in the expansion of the 
averaged Green's function $\hat{g}(\bbox{r},\bbox{p}) = \hat{g}(\bbox{r})
+ \bbox{p\hat{g}}_1(\bbox{r})+\ldots$ over Legendre polynomials. This 
problem can be solved only for $\Gamma\ll 1$ by expanding the right-hand side 
of Eq.\ (3) into a power series in $\Gamma$, which was used in Ref.\ 
\onlinecite{8} for calculating the corrections to the Josephson current of
the order of $\Gamma^2$.

In this paper, we pay attention, first of all, to the fact that the problem of 
calculation of the current--phase relation for a diffusive junction in the 1D 
geometry has the sense only in the case of low transmissivity of the barrier. 
Indeed, simple estimates obtained on the basis of the well-known formula for 
$j(\Phi)$ in the first order in $\Gamma$, 
\begin{equation}
j_0(\Phi) = I(\Delta)\tanh(\Delta/ 2T)\sin\Phi
\end{equation}
(which coincides, according to the Anderson theorem, with the 
Ambegaokar--Baratoff result for a pure superconductor \cite{9}), show that 
{\em even for small $\Gamma \sim l/ \xi_0 \ll 1$ the critical current through 
the junction becomes of the order of the bulk thermodynamic critical current} 
$n_sev_{sc}$, where $v_{sc} \sim 1/m\xi_0$ is the critical velocity of the 
condensate, $n_s \sim m\nu_FD\Delta$ its density, $m$ the electron mass  
($\hbar = 1$). Thus, for $\Gamma \gg l/\xi_0$ the tunnel junction does not 
play any longer the role of ``weak link'' with the jump of the order parameter 
phase $\Phi$ and other features of a Josephson element. This follows even 
from the boundary conditions Eq.\ (3) if we use the estimate $\nabla\hat{g} 
\sim \hat{g}/ \xi_0$ in the vicinity of the junction, which leads to 
$[\hat{g}_L, \hat{g}_R] \sim \sin\Phi \sim \xi_0/l\Gamma \ll 1$ 
for $\Gamma \gg l/ \xi_0$ \cite{note3}. This 
criterion of weak link can be also formulated in terms of the conductance of 
the system in the normal state: {\em the resistance of the junction must 
exceed the resistance of a metal layer of thickness $\xi_0$}.

From this it follows that the parameter
\begin{equation}
W = (3\xi_0/4l) \Gamma \gg \Gamma
\end{equation}
plays a fundamental role in the theory of Josephson effect for diffusive 
junction (the factor 3/4 is chosen for convenience of notation). We can 
attach to this parameter the meaning of the effective tunneling probability 
for Cooper pairs, which is higher than the conventional probability $\Gamma$ 
of quasiparticle tunneling. Small values of $W \ll 1$ correspond to ``weak 
link'' conditions (Josephson effect); for $W > 1$, the presence of a tunnel 
barrier virtually does not affect the supercurrent flow and the distribution 
of the order parameter in a diffusive superconductor. Moreover, we can expect 
that {\em just $W$ and not $\Gamma$ is a true parameter of the expansion 
of $j(\Phi)$ in the barrier transmissivity}. Indeed,  
the dependence of the Josephson current on the mean free path is absent
only within the main approximation in $\Gamma$, Eq.\ (4) and, therefore, 
it must be manifested in higher-order terms of the expansion of 
$j(\Phi)$ in the emergence of additional dimensionless parameter $\xi_0/l$ 
in them, which vanishes at $l \rightarrow \infty$. An analysis of corrections 
to the current--phase dependence of Eq.\ (4), carried out in Sec.\ 4 of this 
article in the next order in $W$, confirms these considerations and proves 
that the corrections $\sim \Gamma^2$ to the Josephson current obtained in 
Ref. \onlinecite{8} and associated with the corrections to boundary conditions 
Eq.\ (3), are much smaller and insignificant in fact.

Another important result of the analysis of the current-carrying state of a 
diffusive Josephson junction is the conclusion concerning the emergence of 
localized states of electron excitations in the vicinity of the barrier. This 
phenomenon is well known for a ballistic tunnel junction \cite{10,11} in which 
discrete energy levels 
\begin{equation}
\epsilon_n(\Phi) = \pm \Delta (1-d\sin^2\Phi/2)^{1/2},
\end{equation}
associated with Andreev localization of electron excitations near the jump in 
the order parameter phase, split from the continuous spectrum in the 
current-carrying state. A similar phenomenon also takes place in a diffusive 
junction in which, however, isolated coherent energy levels cannot exist due 
to electron scattering at impurities and defects. In this case, the most 
adequate description of the variation of the energy spectrum of excitations 
is the deformation of their local density of states $N(\epsilon,\bbox{r})=
\;\mbox{Re}\; u^R(\epsilon,\bbox{r})$ ($u^R$ is the diagonal component of the 
retarded Green's function for the superconductor) which is assumed for 
brevity to be normalized to its value $\nu_F$ in the normal metal. In the 
absence of current, the density of states in a homogeneous superconductor has 
the standard form $N_0(\epsilon)=|\epsilon|\Theta(\epsilon^2-\Delta^2)/
\sqrt{\epsilon^2-\Delta^2}$ ($\Theta(x)$ is the Heaviside function) with root 
singularities at the gap boundaries. In the current state, the momentum $p_s$ 
of the superfluid condensate plays the role of a depairing factor smoothing 
the singularities of $N(\epsilon)$ and reducing the energy gap $2\epsilon_\ast$ 
by $\Delta-\epsilon_\ast(p_s)\sim(Dp_s^2)^{2/3}$ \cite{12}. In the vicinity of 
a weak link, a similar (and main) factor of the energy gap suppression is the 
phase jump $\Phi$ which leads to the formation of {\em a ``potential well'' 
around the junction having a width of the order of $\xi_0$ and containing 
localized excitations with an energy $|\epsilon| < \Delta$} (see Sec.\ 3).
In contrast to the ballistic case, {\em the Josephson transport in a diffusive 
junction is performed not only by the states in the potential well, but by 
excitations within the whole energy region near the gap edge where the 
density of states differs significantly from the unperturbed value}.

\section{Equations for Green's function of a low-transparent Josephson 
junction}

In order to calculate the density of states and equilibrium supercurrent
\begin{equation}
j = {e \over 4} \nu_F v_F D \int^{+\infty}_{-\infty} \!\!\!\! d\epsilon 
f_0(\epsilon)\mbox{Tr} \sigma_z (\hat{g}^R \nabla \hat{g}^R\!
  -\!\hat{g}^A \nabla \hat{g}^A)(\epsilon)
\end{equation}
we must solve equations for the matrix retarded (advanced) Green's functions 
$\hat{g}^{R,A}(\bbox{r},\epsilon)$ averaged over the ensemble of scatterers:
\begin{equation}
[\sigma_z \epsilon +\Delta \exp (i\sigma_z\chi) i \sigma_y, \hat{g}]=
iD \nabla (\hat{g} \nabla \hat{g}), 
\end{equation}
Here $\Delta$ and $\chi$ are the modulus and phase of the order parameter and 
$f_0(\epsilon)=(1/2)(1+\tanh(\epsilon/2T))$ is the equilibrium distribution 
function.

According to the normalization condition $\hat{g}^2=1$ for the Green's 
function, the matrix $\hat{g}$ can be presented as $\hat{g} = 
\bbox{\sigma u}$, where $\bbox{\sigma}$ is the vector formed by Pauli 
matrices. Using the well-known relations $(\bbox{\sigma a})(\bbox{\sigma b})
=\bbox{ab} + i\bbox{\sigma} [\bbox{a} \times \bbox{b}]$, $[\sigma_z, 
\bbox{\sigma}] = 2i[\bbox{\sigma} \times \bbox{s}]$, where $\bbox{s}$ is the 
unit vector of ``isotopic spin'' directed along the $z$-axis in the space of 
Pauli matrices, we can obtain from Eqs.\ (3) and (8) the following equations 
and the boundary conditions for the vector Green's function $\bbox{u}$:
\begin{equation}
\epsilon[\bbox{s} \times \bbox{u}]+i\Delta [\bbox{\chi} \times \bbox{u}] = 
(D/2) \nabla [\bbox{u} \times \nabla \bbox{u}],\;\; \bbox{u}^2 = 1,
\end{equation}
\begin{equation}
\xi_0 [\bbox{u} \times \nabla \bbox{u}]_{L,R} = 
2W [\bbox{u}_L \times \bbox{u}_R],
\end{equation}
where $\bbox{\chi} = (\sin \chi, \cos\chi,0)$ is the symbolic vector of the 
order parameter phase.

Singling out the component of the vector $\bbox{u}$ along the direction 
$\bbox{s}$: $\bbox{u}=\bbox{s}u+i\bbox{v}$ ($\bbox{vs} =0$), we project 
Eq.\ (9) onto the $(x,y)$-plane in the space of Pauli matrices:
\begin{equation}
\epsilon \bbox{v} - \Delta u \bbox{\chi} = (iD/2) \nabla (u \nabla \bbox{v} - 
\bbox{v} \nabla u), \;\; u^2 - \bbox{v}^2 = 1
\end{equation}
and introduce the unit vector $\bbox{\psi} = (\sin\psi,\cos\psi,0)$ directed 
along $\bbox{v}$: $\bbox{v} = \bbox{\psi} v$, where $\psi(\bbox{r},\epsilon)$
is the phase of ``anomalous'' Green's function $v$ ($\nabla\bbox{\psi} = 
[\bbox{\psi}\times\bbox{s}]\nabla\psi$). The obtained system of scalar 
equations is a possible representation of Usadell equations:
\begin{equation}
\epsilon v\!-\!\Delta u\cos(\psi\!-\!\chi)\!=
\!{i \over 2} D \left[\nabla(u \nabla \!v\!-\!v\nabla\! u)\!-\!
uv(\nabla\!\psi)^2\right]\!,
\end{equation}
\begin{equation}
\Delta v \sin(\psi-\chi) = (iD/ 2)\nabla(v^2 \nabla \psi),
\end{equation}
\begin{equation}
u^2 - v^2 = 1,
\end{equation}
and its solutions determine the supercurrent
\begin{equation}
j(\Phi) = -e \nu_F D\int_{-\infty}^{+\infty} d\epsilon f_0 
\;\mbox{Im}\; (v^R)^2 \nabla \psi^R.
\end{equation}

Choosing the coordinate axis $x$ orthogonally to the contact plane $x = 0$ 
($\chi(+0) = -\chi(-0) = \Phi/2$) and taking into account the continuity of 
Green's function and antisymmetry of their derivatives, we can easily obtain 
from Eq.\ (10) the boundary conditions to Eqs.\ (12), (13) for $x \rightarrow 
+0$:
\begin{equation}
\xi_0(u\nabla v-v\nabla u)(0) = 4Wu(0)v(0)\sin^2\psi(0),
\end{equation}
\begin{equation}
\xi_0\nabla \psi(0) = 2W\sin 2\psi(0).
\end{equation}

Far away from the junction, the behavior of the order parameter and Green's 
function phases is described by linear asymptotic form corresponding to 
the given value of current 
\begin{equation}
\chi(+\infty)\! =\! \psi(+\infty)\! =\! \chi_\infty+2p_s x, \;\;  
p_s\! =\! (W/\xi_0)\sin\Phi,
\end{equation}
i.e., of the superfluid momentum $p_s$ whose magnitude is determined in the 
main approximation by the condition of equality of the current Eq.\ (4) 
through the junction to its value $j = \pi e\nu_F Dp_s \Delta 
\tanh(\Delta/2T)$ in the bulk of the metal. The Green's functions tend to 
their asymptotic values satisfying Eqs.\ (12)--(14) for $\psi = \chi$ and 
$\nabla u = \nabla v = 0$.

Using the parametrization $u = \cosh\theta, v = \sinh\theta$, which takes into 
account the normalization condition Eq.\ (14), we can put in correspondence to 
the vector Green's function $\bbox{u}$ the following geometrical image 
\cite{13}. The unit vector $\bbox{u}$ in a normal metal is directed along the 
isospin axis $z$ (which corresponds to a purely electron or hole state of 
excitation of a Fermi gas), while in a superconductor this vector is deflected 
from the axis through an imaginary angle $i\theta$  and turned around it 
through the azimuthal angle $\psi$. In the spatially homogeneous case, this 
angle obviously coincides with the phase of the order parameter 
($\psi = \chi$), and the scalar Green's functions $u$ and $v$ are described by 
the formulas
\begin{equation}
u^{R,A}\! =\! \cosh\theta_s\! =\! {\epsilon \over \sqrt{(\epsilon\pm i0)^2-
\Delta^2}}, \;\; v^{R,A}\! =\! \sinh\theta_s,	
\end{equation}
%
where $\pm i0$ defines the position of singularities of the retarded 
(advanced) Green's function in the complex plane $\epsilon$, and the square 
root in Eq.\ (20) is defined so that $u^{R,A} \rightarrow \pm 1$ for 
$\epsilon \rightarrow +\infty$.

Eqs.\ (12)--(14) for Green's functions should be supplemented by the 
self-consistency conditions for the modulus and phase of the order parameter:
\begin{equation}
\Delta = \lambda\int^{+\infty}_{-\infty} d\epsilon f_0 \;\mbox{Re}\; v^R,
\end{equation}
\begin{equation}
\int_{-\infty}^{+\infty} d\epsilon f_0 \;\mbox{Re}\; v^R\sin(\psi^R-\chi)=0,
\end{equation}
where $\lambda$ is the constant of superconducting interaction. Taking into 
account the current conservation law, Eqs.\ (13) and (21), it is convenient 
to calculate the value of current at the barrier ($x \rightarrow +0$) by 
expressing $\nabla\psi(0)$ in Eq.\ (15) with the help of Eq.\ (17) through 
the phase jump $2\psi(0)$: 
\begin{equation}
j(\Phi)\! =\! -{e\over 2}\nu_F v_F \Gamma \!\int_{-\infty}^{+\infty}\!\!\! 
d\epsilon f_0 \;\mbox{Im}\; (v^R(0))^2\sin 2\psi^R(0),
\end{equation}
which allows us to single out explicitly the small parameter of the theory, 
i.e., the barrier transmissivity $\Gamma$. It can easily be verified that in 
the main approximation using the unperturbed values of Green's function of
Eq.\ (19) and phase $\psi(0) \approx \chi(0) = \Phi/2$, Eq.\ (22) leads to 
the result of Eq.\ (4).

A simplifying factor in the case of a low transmissivity of the barrier is 
that the quantities $\psi - \chi$ and $\nabla \psi$ proportional to the 
current through the junction are small (see Eqs.\ (17) and (13)), and hence we 
can omit in Eq.\ (12) the terms quadratic in $W$ and containing the phase
gradients. Replacing $\psi(0) \approx \chi(0) = \Phi/2$ in the boundary 
conditions Eqs.\ (16), (17), to the same degree of accuracy, we obtain the 
equation and the boundary conditions for the parameter $\theta$:
\begin{equation}
\epsilon\sinh\theta - \Delta(x)\cosh\theta = (iD/2)\nabla^2\theta,
\end{equation}
\begin{equation}
\xi_0\nabla\theta(0) = 2W\sinh 2\theta(0)\sin^2\Phi/2, \;\; 
\theta(+\infty)=\theta_s.
\end{equation}

Direct application of the perturbation theory to the solution of Eq.\ (23) 
($\theta(x)=\theta_0+\theta_1(x)$, $\Delta(x)=\Delta_0+\Delta_1(x)$) leads to 
an expression for the correction $\theta_1(x)$ containing nonintegrable 
singularities at the gap boundaries, and as a consequence, to the divergence 
of the corresponding correction to the Josephson current Eq.\ (4). This is 
associated with the emergence of localized states of quasiparticles at a 
tunnel junction in the current-carrying state mentioned in Introduction 
and considered in the next section.

\section{Localized states at a tunnel barrier}

It will be proved below that the depth of the ``potential well'' in the 
vicinity of the barrier is much larger than the scale of variation of the 
order parameter. Consequently, it is sufficient to confine an analysis of the 
behavior of the density of states to the model with a constant $\Delta$, in 
which Eq.\ (23) has a simple solution describing the attenuation of 
perturbations of Green's functions at a distance $\sim \xi_0$ from the 
barrier:
\begin{mathletters}
\begin{equation}
\tanh{\theta(x)-\theta_s \over 4}=\tanh{\theta(0)-\theta_s \over 4}
\exp(-k_\epsilon|x|), 
\end{equation}
\begin{equation}
k^{-2}_\epsilon = i\xi_0^2\sinh\theta_s, \; \;\mbox{Re}\; k_\epsilon > 0.
\end{equation}
\end{mathletters}

The quantity $\theta(0)$ satisfies the boundary condition following from 
Eqs.\ (24) and (25):
\begin{equation}
k_\epsilon\xi_0\sinh{\theta_s\!-\!\theta(0)\over 2}\! =\! 
\gamma\sinh 2\theta(0), \; \gamma\! =\! W\sin^2{\Phi\over 2}\! \ll\! 1,
\end{equation}
which can be reduced to the eighth-power algebraic equation in $z =
\exp \theta(0)$:
\begin{equation}
2z^3(z-z_s)^2 = i\gamma^2(z_s^2-1)(z^4-1)^2, \;\; z_s = \exp\theta_s.
\end{equation}

In the general case (for an arbitrary $\epsilon$), the solution of Eq.\ (27) 
can be obtained only numerically, but the presence of the small parameter 
$\gamma$ in (26) and (27) makes it possible to apply the perturbation theory. 
Far away from the spectrum boundary, we can put $\theta(0) = \theta_s$ on 
right-hand side of (26), which leads to the following expression for the 
correction to the density of states at the barrier:
\begin{equation}
N(\epsilon,0)-N_0(\epsilon)=
-2\gamma\;\mbox{Re}\left(\sqrt{i\sinh^3\theta_s}\sinh 2\theta_s\right),
\end{equation}
that becomes obviously inapplicable for $|\epsilon| \rightarrow \Delta$, 
where $|\theta_s| \rightarrow \infty$. In this region, we must apply the 
improved perturbation theory (IPT) by putting $|z|, |z_s| \gg 1$ for an 
arbitrary (not necessarily small) value of $z - z_s$. This not only reduces 
the power of the general Eq.\ (27), but also allows us to write it in a 
universal form which does not contain the depairing parameter $\gamma$:
\begin{mathletters}
\begin{equation}
(y\sqrt{E}-1)^2=iy^5,
\end{equation}
\begin{equation}
y\!=\!z/\beta\sqrt{2},\; E\!=\!\beta^2(\epsilon\!-\!\Delta)/\Delta,\; 
\beta\!=\!(2/\gamma)^{1/5} \gg 1,
\end{equation}
\end{mathletters}

Relations Eq.\ (29) show that the increase in the density of states is bounded 
by a quantity of the order of $\beta \sim W^{-2/5}$ as we approach the 
spectrum boundary. Thus, the range of applicability of the conventional 
perturbation theory, Eq.\ (28), is determined by the condition $(\epsilon - 
\Delta)/ \Delta \gg \beta^{-2}$ and overlaps with the region of 
applicability $(\epsilon - \Delta)/ \Delta \ll 1$ of the IPT. The boundary 
$\epsilon_\ast$ of the spectrum (the position of the bottom of the potential 
well), below which the density of states vanishes, corresponds to the 
emergence of a purely imaginary root of Eq.\ (29a) at the point $E_\ast = 
-(25/6)(2/3)^{1/5} \approx -3.842$:
\begin{equation}
\epsilon_\ast(\Phi)\! =\! \Delta\!\!\left[\!1\!-\!C\!\!\left(\!\!
W\sin^2\!{\Phi\over 2}\!\right)^{\!4/5}\right]\!\!, 
\;
C\!=\!{25 \over 3\!\cdot\! 6^{1/5}}\! \approx\! 5.824.
\end{equation}

\begin{figure}[tb]
\epsfxsize=7.5cm\centerline{\epsffile{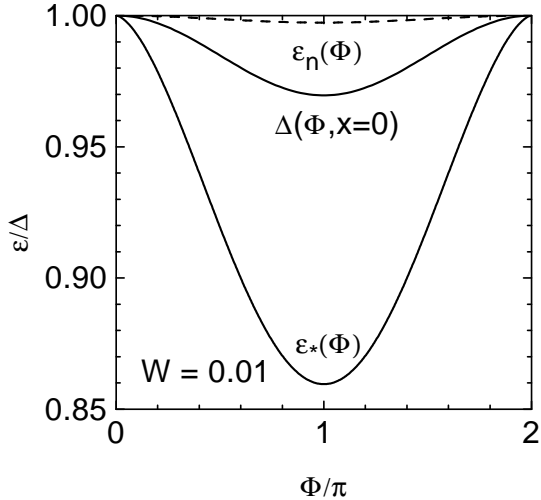}}
\vspace{0.1in}
\caption{Phase dependence of the position of the bottom of the ``potential 
well'' $\epsilon_\ast$, Eq.\ (30), and the order parameter $\Delta(0)$, 
Eq.\ (48), in the vicinity of the tunnel junction (solid curves) at $T = 0, 
W = 0.01$ and $\xi_0/l = 5$. The dashed curve shows for comparison the 
position of the Andreev level in a pure single-mode junction, Eq.\ (5), for the 
same barrier transmissivity.}
\label{fig1}
\end{figure}

\begin{figure}[tb]
\epsfxsize=7.5cm\centerline{\epsffile{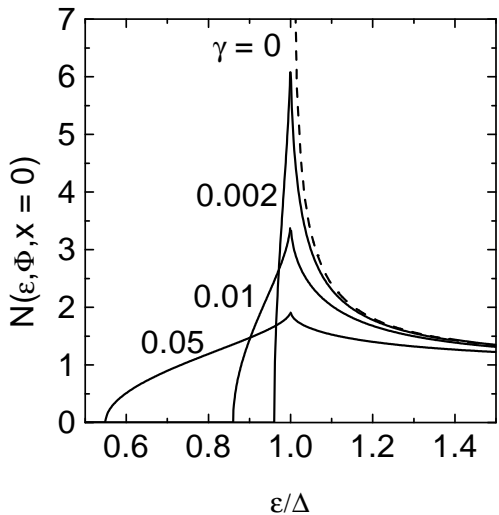}}
\vspace{0.1in}
\caption{Dependence of the density of states $N(\epsilon,\Phi,0)$ at the 
tunnel junction on the energy of quasiparticles for various values of the 
depairing parameter $\gamma = W\sin^2(\Phi/2)$ (solid curves). The dashed 
curve shows the energy dependence of the unperturbed density of states 
$N_0(\epsilon)$.}
\label{fig2}
\end{figure}

The dependence of the position of the spectrum boundary on the phase jump at 
the junction is illustrated by Fig.\ \ref{fig1} in which a similar dependence 
of the position of the Andreev level Eq.\ (6) in a junction between pure 
superconductors is shown for comparison. It should be noted that the scale of 
variation of $\epsilon_\ast(\Phi)$ is much larger than the splitting of the 
Andreev level from the boundary of the continuous spectrum for the same 
barrier transmissivity. This is associated with the large value of the
depairing parameter $\gamma$ in the diffusive junction as compared to the 
splitting parameter $\Gamma$ of the Andreev level as well as with the large 
numerical value of the constant $C$ defining the shift of the spectrum 
boundary Eq.\ (30). Fig.\ \ref{fig2} shows the results of numerical 
calculation of the density of states at the junction on the basis of the 
general formula Eq.\ (27) for different values of the depairing parameter, 
which show that in addition of the root singularity ($\sim \sqrt{\epsilon - 
\epsilon_\ast}$) at the spectrum boundary, the quantity $N(\epsilon)$ has a 
``beak-type'' root singularity for $\epsilon = \Delta$. Its physical nature is 
associated with an infinite increase in the attenuation length 
$k_\epsilon^{-1}$ of the perturbation of Green's function in the bulk of the 
metal, Eq.\ (25), within the vicinity of the gap boundary.

\begin{figure}[tb]
\epsfxsize=7.5cm\centerline{\epsffile{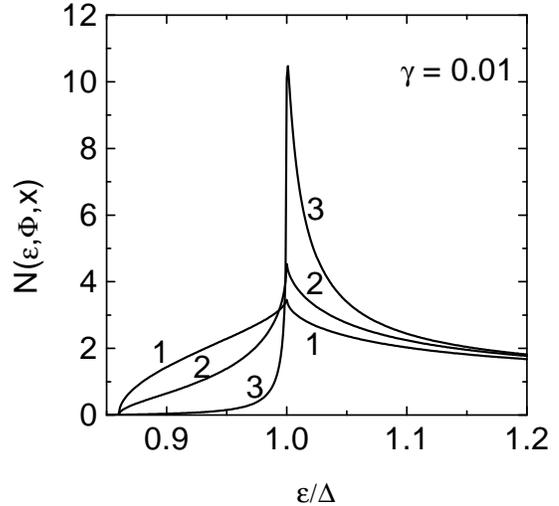}}
\vspace{0.1in}
\caption[]{Energy dependence of the density of states $N(\epsilon,\Phi,0)$ 
for $\gamma = 0.01$ at different distances $x$ from a diffusive tunnel 
junction: 0 (curve {\em 1}), $\xi_0$ (curve {\em 2}), and $5\xi_0$ 
(curve {\em 3}).}
\label{fig3}
\end{figure}

For $\epsilon_\ast < \epsilon < \Delta$, the density of states decreases 
exponentially with increasing distance from the junction (Fig.\ \ref{fig3}), 
which corresponds qualitatively to the image of the potential well of depth 
$\Delta - \epsilon_\ast$ and of width $\sim \xi_0$ with excitations localized 
in it.

\begin{figure}[tb]
\epsfxsize=7.5cm\centerline{\epsffile{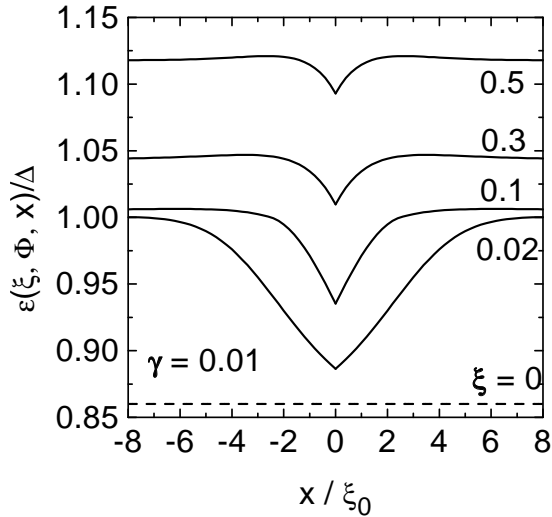}}
\vspace{0.1in}
\caption{Lines corresponding to the number of states of quasiparticles 
$\xi(\epsilon,\Phi,x) = {\rm const}$ (Eq.\ (33)) for $\gamma = 0.01$ in the 
vicinity of the junction. The dashed line shows the position of the bottom of 
the ``potential well'' ($\xi = 0, \epsilon = \epsilon_\ast(\Phi)$).}
\label{fig4}
\end{figure}

It is well known that the Josephson current is carried through a ballistic 
junction by localized excitations only and can be presented in the following 
form:
\begin{equation}
j(\Phi) = -2e\sum_n{\partial\epsilon_n(\Phi)\over \partial\Phi}
\tanh{\epsilon_n(\Phi)\over 2T},
\end{equation}
where the index $n$ labels Andreev levels. At the same time, Eq.\ (22) for 
current expressed in the IPT approximation in terms of the reduced variables 
of Eq.\ (29),  
\begin{equation}
j(\Phi)\!\approx\! -I(\Delta)\tanh{\Delta\over 2T}\sin\Phi\!\!\!\!\int\limits
_{E_\ast(\Phi)}^\infty\!\!\!\!{dE \over \pi}\mbox{Im}\left(y^R\right)^2\!\! 
=\! j_0(\Phi),
\end{equation}
shows that the charge transfer in a diffusive junction is performed not only 
by the states within the potential well ($E < 0, \epsilon < \Delta$), but also 
by the excitations with energy $\epsilon > \Delta$ in the region $\epsilon - 
\Delta\sim\Delta\beta^{-2}$, where the density of states differs significantly 
from the unperturbed value $N_0(\epsilon)$. It should be noted in this 
connection that Argaman \cite{14} proposed an analog of Eq.\ (31) for a 
diffusive system, which can be obtained by the replacement of the energy 
$\epsilon_n(\Phi)$ of Andreev levels by the local value $\epsilon(\xi,\Phi,x)$ 
of the excitation energy for $x = 0$, which is adiabatically deformed by 
supercurrent, using instead of the discrete number $n$ the continuous variable
\begin{equation}
\xi = \int_{\textstyle{\epsilon_\ast(\Phi)}}^{\textstyle{\epsilon(\xi,\Phi,x)}} 
d\epsilon^\prime N(\epsilon^\prime,\Phi,x)
\end{equation}
viz., the number of states with an energy smaller than $\epsilon$ ($\xi = 
\Theta(\epsilon^2-\Delta^2)\sqrt{\epsilon^2-\Delta^2}$ for a homogeneous 
superconductor) \cite{note4}. One can assume that the contributions from the 
bound and delocalized states to the Josephson current are taken into account 
simultaneously by the formula
\begin{equation}
j(\Phi)=-2e\nu_F\int_0^\infty d\xi\; 
{\partial\epsilon(\xi,\Phi,0)\over \partial\Phi}
\tanh{\epsilon(\xi,\Phi,0)\over 2T},
\end{equation}
which, however, leads to correct results only in the case of a homogeneous 
current-carrying state (where $\nabla\chi$ plays the role of $\Phi$) or a 
wide $SNS$-junction (with a width $L \gg \xi_0$ of the normal layer) and is 
inapplicable for a narrow bridge and tunnel junction. Nevertheless, the 
consideration of the function $\epsilon(\xi,\Phi,x)$ is useful in these cases 
also since this allows us to visualize the variation of the energy 
distribution of quasiparticle states in the vicinity of the junction
(Fig.\ \ref{fig4}).

\section{Current--phase dependence for a junction in the second order in $W$}

Although the modified perturbation theory for Green's function in the energy 
representation described in the preceding section is the most physically 
obvious method operating with actual excitation energies, it leads to 
considerable formal difficulties in the calculation of corrections to the 
Josephson current Eq.\ (4). Indeed, it was shown in the previous section that 
the expression for $j(\Phi)$ calculated on the basis of the IPT for Green's 
functions, Eq.\ (32), coincides with Eq.\ (4) since the small IPT parameter 
$\beta^{-2}$ cancels out as we go over to the reduced variables of Eq.\ (29). 
Thus, in order to calculate the corrections to Eq.\ (4) we are interested in, 
we must leave the approximation of Eq.\ (29) that describes the behavior of 
Green's functions correctly only in a narrow vicinity of singularity in the 
density of states. For this purpose, it is convenient to use the formalism of 
temperature Green's functions by going over from integration over energy in 
Eqs.\ (20)--(22) to summation over the Matsubara frequencies $\omega_n = \pi 
T(2n + 1), n = 0, \pm 1, \pm 2, \ldots$:
\begin{equation}
j(\Phi) = -\pi e\nu_F v_F \Gamma T\sum_{\omega_n>0}\; \mbox{Re}\; 
v^2(0)\sin 2\psi(0),
\end{equation}
\begin{equation}
\Delta(x) = -2\pi\lambda T\sum_{\omega_n>0} \;\mbox{Im}\; v(x)
\end{equation}
and making the substitution $\epsilon \rightarrow i\omega_n$ in Eq.\ (23). 
This allows us to avoid divergences of the type of Eq.\ (28) in the 
perturbation theory which, unlike the IPT, makes it possible to take into 
account the coordinate dependence $\Delta(x)$.

It is expedient to use as the main approximation in the asymptotic expansion 
$\theta = \theta_0 + \theta_1 + \ldots$ the ``adiabatic'' value of Green's 
function corresponding to the local value of $\Delta(x)=\Delta+\Delta_1(x)$, 
($\Delta_1(\infty)=0$):
\begin{equation}
u_0(\!x\!)\! =\! \cosh\theta_0(\!x\!)\! =
\! {{\omega_n}\over{\tilde{\omega}_n}(\!x\!)},\; 
v_0(\!x\!)\! = \!\sinh\theta_0(\!x\!)\! =
\!{\Delta\over{\tilde{\omega}_n}(\!x\!)}, 
\end{equation}
%
where ${\tilde{\omega}_n}(x) = \sqrt{{\omega_n}^2+\Delta^2(x)}$. In this case, 
the correction $\theta_1(x)$ satisfies the nonhomogeneous equation
\begin{equation}
\nabla^2\theta_1 -k^2_\omega \theta_1 = \nabla^2\theta_0, \quad 
k^2_\omega = 2{\tilde{\omega}_n}/ D
\end{equation}
with the boundary conditions $\nabla\theta_1(+0)=2W\sinh 2\theta_s \times 
\sin^2\Phi/2$, $\theta_1(\infty)=0$, where $\cosh\theta_s = {\omega_n}/
{\tilde{\omega}_n}$ is the value of the Green's function far away from the 
junction with the unperturbed value of $\Delta$, and ${\tilde{\omega}_n} =
\sqrt{{\omega_n}^2+\Delta^2}$.

The self-consistency condition for $\Delta_1(x)$ following from Eq.\ (20), 
\begin{equation}
\Delta_1(q)T \sum_{\omega_n > 0} {\Delta^2 \over {\tilde{\omega}_n}^3}=
-T\sum_{\omega_n>0}{{\omega_n}\over{\tilde{\omega}_n}}\;\mbox{Im}\;
\theta_1(i\omega_n,q)
\end{equation}
completes the system of equations for determining the corrections $\theta_1$ 
and $\Delta_1$, whose solution in the Fourier representation has the form
\begin{mathletters}
\begin{equation}
\Delta_1(q) = -8W \Delta {B(q)\over \xi_0 A(q)} \sin^2 \Phi/2, 
\end{equation}
\begin{equation}
\theta_1(i\omega_n,q) = 8W\Delta {i\omega_n \over {\tilde{\omega}_n}}
{1 \over q^2+k^2_\omega} {A(0) \over \xi_0 A(q)} \sin^2 \Phi/2,
\end{equation}
\end{mathletters}
\begin{mathletters}
\begin{equation}
A(q) = A(0)+q^2B(q),\;\; 
A(0)=2\pi T\sum_{\omega_n>0} {\Delta^2 \over {\tilde{\omega}_n}^3}, 
\end{equation}
\begin{equation}
B(q)=2\pi T \sum_{\omega_n>0} {\omega_n^2 \over {\tilde{\omega}_n}^3}
{1 \over q^2+k^2_\omega}.
\end{equation}
\end{mathletters}
\[
\left(\theta_1(i\omega_n,x),\Delta_1(x)\right)\! =\!\!\!
\int_{-\infty}^{+\infty}\!\!{dq \over 2\pi} e^{iqx}\! 
\left(\theta_1(i\omega_n, q), \Delta_1(q)\right).
\]

As regards the correction to the asymptotic value Eq.\ (18) of the phase 
$\psi(x)$ of the Green's function, it is equal to zero in this approximation. 
In order to prove this, we introduce the quantity $\varphi = \psi - \chi \ll 
1$, which, according to Eq.\ (13), obeys the equation
\begin{equation}
\nabla^2\varphi - k^2_\omega \varphi = -\nabla^2\chi_1,
\end{equation}
where $\chi_1 = \chi(x) - \chi(\infty)$ is a correction to Eq.\ (18) localized 
near the junction. Taking into account the boundary condition $\nabla 
\varphi(0) = -\nabla\chi_1(0)$ following from Eqs.\ (17) and (18), we find 
that this equation has the simple solution $\varphi(i\omega_n,q)=
-q^2\chi_1(q)/(q^2+k_\omega^2)$ which leads, after the substitution into the 
self-consistency condition Eq.\ (21), to the homogeneous integral 
equation for $\chi_1(q)$:
\begin{equation}
T\sum_{\omega_n>0}{\Delta \over {\tilde{\omega}_n}} \int_{-\infty}^{+\infty} 
dq {q^2 \cos qx \over q^2+k_\omega^2} \chi_1(q) = 0.
\end{equation}

The only nonsingular solution of Eq.\ (43) is $\chi_1(q) \equiv 0$, which 
proves the absence of a correction to the Josephson current due to the 
deviation of the behavior of the phases of the order parameter and Green's 
functions from the linear law Eq.\ (18). This result can be explained as 
follows. The correction $\chi_1(x)$ is obviously of the order of the small
correction $p_{s1}(x)$ to the constant value $p_s$ of Eq.\ (18) in the 
vicinity of the junction, that ensures the conservation of the current upon 
a change in $N(\epsilon)$ and $\Delta$. Since the value of $p_s \sim W$, the 
correction to this quantity, and hence $\chi_1(x)$ and $\varphi$ have a 
higher order of smallness ($\sim W^2$) than the corrections of the order of 
$W$ we are interested in.

Substituting Eqs.\ (40), (41) into Eq.\ (22), we obtain the required 
correction to the Josephson current:
\[
\delta j\!=\!j(\Phi)\!-\!j_0(\Phi)\!=
\!-{4T \over \Delta} I(\Delta) \sin\Phi \sum_{\omega_n > 0} \mbox{Re}
\left(v^2\! + \!{\Delta^2 \over {\tilde{\omega}_n}^2} \right)\!=
\]
\begin{equation}
= -I(\Delta) W_0 Z(T)\left(\sin\Phi-{1 \over 2}\sin 2 \Phi \right),
\end{equation}
\begin{equation}
Z(T) \! = \! {16 \over \pi} \sqrt{\Delta\Delta_0} T \!\!
\sum_{\omega_n > 0} \! {\omega_n^2 \over {\tilde{\omega}_n}^4}
\! \int\limits^{+\infty}_{-\infty} \!\! {dk \over k^2\!+
\! \tilde{k}^2_\omega} \!\left[\!1\! + \!{\tilde{k}^2_\omega B(k) \over A(k)}
\!\right] \!\!, 
\end{equation}
%
where $\tilde{k}_\omega = {\tilde{\omega}_n}/\Delta$, $A(k)$ and $B(k)$ are 
defined by Eqs.\ (41) upon the substitution $k_\omega \rightarrow 
\tilde{k}_\omega$, and $W_0$ and $\Delta_0$ are the values of $W$ and $\Delta$ 
at $T = 0$.

At low temperatures ($T \ll \Delta$), the summation over $\omega_n$ in 
Eqs.\ (41) and (45) can be replaced by integration with respect to the 
continuous variable $\omega$:
\[
A(0)=1, \;\; B(k)=\int_0^\infty{\tanh^2 v\; dv \over k^2+\cosh v}=
\]
\[
={1 \over k^4}\left({\pi \over 2}
-2\sqrt{1-k^2}\arctan\sqrt{1-k^2 \over 1+k^2} - k^2\right),
\]
which leads to the following asymptotic value of the function $Z(T)$ for 
$T \rightarrow 0$:
\begin{equation}
Z(T)\! =\! {8\over \pi^2}\!\! \int_0^\infty\!\!\!\! dk\! 
\left[{\pi k^2 \over (1\!+\!k^2)^{9/4}}\!+\!
{2B^2(k) \over 1\!+\!k^2B(k)}\right]\!\approx\!2.178.
\end{equation}
%

\begin{figure}[tb]
\epsfxsize=7.5cm\centerline{\epsffile{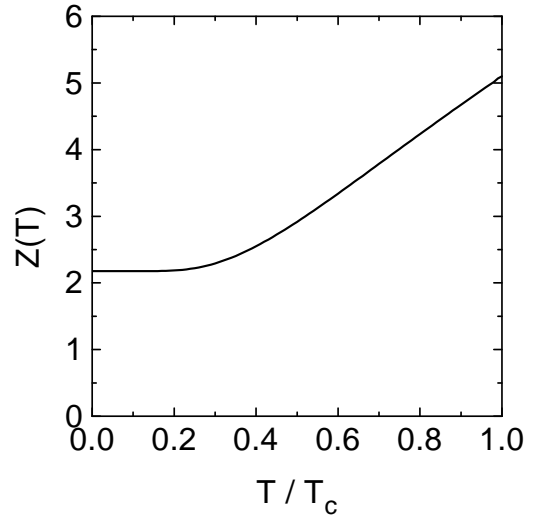}}
\vspace{0.1in}
\caption{The function $Z(T)$, Eq.\ (46), defining the temperature dependence 
of the ratio $\delta j(\Phi)/I(\Delta)$, Eq.\ (44).}
\label{fig5}
\end{figure}

In the vicinity of critical temperature ($\Delta \ll T$), the quantity $A(0)
\approx 7\zeta(3)\Delta^2/4\pi^2 T^2$ is small, and the main contribution to 
integral of Eq.\ (45) comes from the region of small wave vectors $k \sim 
\Delta/T$ corresponding to damping of perturbations at large distances of the 
order of $\xi(T) \propto (T_c - T)^{-1/2}$. This allows us to replace the 
function $B(k)$ by its value $\pi\Delta/4T$ for $k = 0$:
\[
Z(T) = {32 \sqrt{\Delta\Delta_0} \over \pi^3 T}\sum_{n \geq 0} 
{1 \over (2n+1)^2} \!\int_0^\infty\!\! {B(0) \; dk \over A(0)+k^2B(0)} = 
\]
\begin{equation}
= 2\pi\sqrt{\pi\Delta_0 \over 7\zeta(3)T_c}
\approx 5.099.
\end{equation}
The results of numerical calculations of the $Z(T)$ dependence within the 
entire temperature range $0 < T < T_c$ are presented in Fig.\ \ref{fig5}.

Similarly, we can calculate by using Eqs.\ (40) and (41) the asymptotic values 
of the correction $\Delta_1(0)$ to the unperturbed value of the order 
parameter at the junction:
\begin{equation}
{\Delta_1\over \Delta_0}\!\! =\! -\alpha(T)W_0\sin^2\!{\Phi \over 2},
\; \alpha(0)\! =\! 3.037, \; \alpha(T_c)\! =\! 5.782.
\end{equation}

The dependence of the order parameter $\Delta(0)$ on the phase jump at the 
junction at $T = 0$ presented in Fig.\ 1 shows that the main contribution to 
the energy gap suppression comes from the depairing mechanism considered in 
Sec.\ 3, and the change in the order parameter is smaller than the variation
of $\epsilon_\ast(\Phi)$.

The structure of the phase and temperature dependences of the correction to 
the Josephson current of Eq.\ (44) in a diffusive superconductor virtually 
coincide with expression Eq.\ (1) for a junction between pure metals except 
the following circumstance noted in Introduction: the parameter of the 
expansion of $j(\Phi)$ in the transmissivity of the junction for $l \ll \xi_0$ 
is not the tunneling probability $\Gamma$, but a considerably larger parameter 
$W$, Eq.\ (5). This allows one to observe higher harmonics of the 
current--phase dependence in diffusive tunnel junction with a comparatively 
high resistance. Koops et al. \cite{17} apparently reported on the first 
experimental results in this field.

The theory discussed above describes the current--phase dependence for a 
diffusive Josephson junction in the whole temperature range $0 \leq T < T_c$ 
except a narrow neighborhood of $T_c$ in which $\Delta/T_c \sim W_0$ 
($\Delta/T_c \sim \Gamma$ in a pure superconductor), and the magnitude of 
corrections Eqs.\ (44) and (1) becomes comparable with $j_0(\Phi)$, while the 
correction Eq.\ (48) to $\Delta$ becomes of the order of its unperturbed 
value. This means that in the definition Eq.\ (5) of the parameter $W$ near 
$T_c$ the coherence length $\xi_0(T)$ describing the characteristic scale of 
spatial variations of Green's function and density of states should be 
replaced by the characteristic length $\xi(T)$ of variation of the order 
parameter (healing length) in the Ginzburg--Landau theory, whose order of 
magnitude is the same as $\xi_0$ far away from $T_c$. Taking into account the 
results of calculations of $j(\Phi)$ for a pure superconductor in the vicinity 
of $T_c$ \cite{5}, we can obtain the following interpolation estimate of the 
effective transmissivity $W$ suitable for any temperatures and mean free 
paths:
\begin{equation}
W \sim \Gamma \xi(T) \left( {1 \over l} + {1 \over \xi(0)}\right).
\end{equation}

As we approach $T_c$, the value of $W$ increases infinitely, which is 
accompanied with a decrease in the phase jump for a given external current 
bounded by its critical value. Thus, in the 1D geometry for an arbitrarily 
large normal resistance of the junction, there exists a narrow region near 
$T_c$ in which the phase difference of the order parameter at the junction 
is small up to values of current of the order of the bulk critical current.

The authors are grateful to T.N. Antsygina and V.S. Shumeiko for fruitful 
discussions.

This research was supported by the Foundation for Fundamental Studies at 
the National Academy of Sciences of the Ukraine (grant No.~2.4/136).


\begin{references}
\vspace{-1cm}

\bibitem{1}
N. van der Post, E.T. Peters, I.K. Yanson, and J.M. van Ruitenbeek, Phys. Rev. 
Lett. {\bf 73}, 2611 (1994).

\bibitem{2}
H. Takayanagi, T. Akazaki, and J. Nitta, Phys. Rev. Lett. {\bf 75}, 3533 
(1995).

\bibitem{3}
W. Haberkorn, H. Knauer, and S. Richter, Phys. Status Solidi {\bf 47}, K161 
(1978); A.V. Zaitsev, Sov. Phys.--JETP {\bf 59}, 1015 (1984); G.B. Arnold, 
J. Low Temp. Phys. {\bf 59}, 143 (1985).

\bibitem{note1}
The transverse size of the junction is assumed to be smaller than the 
Josephson penetration depth, which ensures the uniform distribution of the 
current over the cross section of the junction.

\bibitem{4}
T.N. Antsygina and A.V. Svidzinskii, Teor. Mat. Fiz. {\bf 14}, 412 (1973).

\bibitem{note2}
The only exception is the case of temperatures close to critical, when the 
presence of the small parameter $\Delta/T_c$ makes it possible to formulate 
the effective computational algorithm of the solution of this problem 
\cite{5,6}.

\bibitem{5}
V.P. Galaiko, A.V. Svidzinskii, and V.A. Slyusarev, Sov. Phys.--JETP {\bf 29}, 
222 (1969).

\bibitem{6}
E.N. Bratus' and A.V. Svidzinskii, Teor. Mat. Fiz. {\bf 30}, 239 (1977).

\bibitem{7}
M.Yu. Kupriyanov and M.F. Lukichev, Sov. Phys.--JETP {\bf 67}, 1163 (1987).

\bibitem{8}
C.J. Lambert, R.Raimondi, V.Sweeney, and A.F. Volkov, Phys. Rev. {\bf B55}, 
6015 (1997).

\bibitem{9}
V. Ambegaokar and A. Baratoff, Phys. Rev. Lett. {\bf 10}, 486 (1963).

\bibitem{note3}
Strictly speaking, this relation contains the jump in the phase of Green's 
function instead of the jump in the order parameter phase, but these 
quantities virtually coincide for $\Gamma \ll 1$ (see Sec.\ 4).

\bibitem{10}
A. Furusaki and M. Tsukada, Phys. Rev. {\bf B43}, 10164 (1991).

\bibitem{11}
S.V. Kuplevakhskii and I.I. Fal'ko, Sov. J. Low Temp. Phys. {\bf 17}, 501 
(1991).

\bibitem{12}
Yu.N. Ovchinnikov, Sov. Phys.--JETP {\bf 32}, 72 (1971).

\bibitem{13}
Yu.V. Nazarov, Phys. Rev. Lett. {\bf 73}, 1420 (1994).

\bibitem{14}
N. Argaman, cond-mat/9709001 (1997).

\bibitem{note4}
The concept of adiabatic deformation of ``energy levels'' in the continuous 
spectrum of a superconducting diffusive system in the current-carrying state 
and their classification on the basis of the continuous ``quantum number'' 
$\xi$ was introduced for the first time in Ref.\ \cite{15} and systematically 
used in Ref.\ \cite{16}.

\bibitem{15}
V.P. Galaiko, Sov. Phys.--JETP {\bf 37}, 922 (1973).

\bibitem{16}
E.V. Bezuglyi and A.Yu. Azovskii, Sov. J. Low Temp. Phys. {\bf 11}, 691 
(1985).

\bibitem{17}
M.C. Koops, G.V. van Duyneveldt, A.N. Omelyanchouk, and R. de Bruyn Ouboter, 
Czech. J. of Phys. {\bf 46} Suppl., 673 (1996).

\end{references}
\end{document}